\documentclass{ws-p8-50x6-00}

\begin{document}

\title{BARYONS 2002: OUTLOOK}

\author{Wolfram Weise}

\address{ECT$^*$, Villa Tambosi, I-38050 Villazzano (Trento), Italy\\
and\\
Physik-Department, Technische Universit\"at M\"unchen,\\
D-85747 Garching, Germany}


\maketitle

\section{Preamble}
This is a sad moment: the outlook concluding this conference should have
been presented by Nathan~Isgur. It will be difficult to meet his standards.

In his introductory talk at BARYONS '98 in Bonn, Nathan drew a picture of his
vision for "Strong QCD". The zeroth order starting point, he hypothesized, for
building hadrons should be relativistic constituent quarks tied together by
flux-tube gluon dynamics. In a book-keeping strategy organized in terms of
inverse powers of $N_c$, the number of colors, one should then add the
quark-antiquark sea and other $1/N_c$ effects as perturbations. We
were also reminded of the potentially important role played by the $q \bar{q}$
vacuum condensate in connecting valence quarks and potentials on one hand with
the current quarks and gluons of QCD on the other.

All of these key notions have been very visible again at the present
conference. So, have we advanced substantially in our understanding of constituent quarks,
gluonic flux tubes, confinement, spontaneous chiral symmetry breaking and the
QCD vacuum since our last BARYONS meeting? I believe the answer is a forceful
"yes", in view of the significant progress achieved by the joint efforts of
experiments and theory, but many issues still remain to be resolved.

\section{Lattice QCD}
With steadily increasing computational power, "solving" QCD on large Euclidean
lattices has now become an important part of hadron physics. Lattice QCD was
addressed by no less than four plenary speakers (C.~Davies, R.~Edwards,
G.~Schierholz, A.W.~Thomas) who gave impressive surveys of the progress made in
recent years. The gluonic flux tube and its translation into a static confining
potential between (infinitely heavy) color sources has become "reality" on the
lattice. The Y-shaped potential characteristic of static three-quark systems is
predicted by lattice QCD, as well as hybrid baryons with three heavy quarks
coupled to gluonic excitations (reviewed by Ph.~Page). But so far, the dynamics of light quarks is not under control, at
least not in the context of the confinement problem.

New developments have been reported on baryon observables
(masses, electromagnetic and axial form factors, moments of structure
functions) extracted from lattice QCD. Improved actions are designed to reduce
discretization errors. Effects of the finite lattice volume are seriously
addressed. Procedures leading beyond the "quenched" (no-fermion-loops)
approximation are well on their way.

Still, these results are limited to relatively large quark masses. The
typical "light" quark masses manageable on the lattice up to now, are 10--20
times larger than the u- and d-quark masses determined at renormalization
scales around 1 GeV. The corresponding pion masses are well above 0.5 GeV, far
away from physical reality, so that important aspects of chiral
pion dynamics are presumably suppressed.

While lattice calculations using small quark masses close to the chiral limit
are out of reach in the foreseeable future, such huge efforts may not even be
necessary. Low-energy QCD in the light-quark sector is realized in the form of
an effective field theory based on chiral symmetry, and this theory can be used
efficiently to extrapolate between lattice results obtained at higher u- and
d-quark masses and the "real world" of small quark masses. The feasibility of
such extrapolations employing suitable Pad\'{e} approximants (Leinweber, Melnitchouk, et~al.)
or extended versions of chiral perturbation theory (Hemmert et al.) has been
discussed at this conference, with promising results. Once the lattice
calculations will approach pion masses around 300 MeV in the near future, reliable
extrapolations using effective field theory methods have a good chance of
closing the remaining gap between lattice QCD and actual observables.

\section{Constituent Quarks}
What is a constituent quark? Over several decades we have learned to live with
this phenomenological concept without really understanding what it means in
detail. Constituent quarks have been remarkably successful in describing ratios
of baryon magnetic moments and organizing the symmetry breaking patterns seen
in the hadron spectrum. One can think of constituent quarks as
quasi-particles in a way analogous to those introduced in many-body
problems. For example, a good approximation for an interacting electron gas at
moderate densities is in terms of weakly interacting quasi-electrons,
i.~e. electrons with their Coulomb interactions screened by a cloud of
electron-hole excitations. Similarly, constituent quarks might be viewed as
quarks dressed by clouds of quark-antiquark pairs and gluons. In distinction
from the behaviour commonly associated with quasi-particles, however,
constituent quarks do {\it not} interact weakly: they experience color
confinement, and their residual interactions must be strong enough to generate
the observed large hyperfine splittings in the baryon spectrum.

One can address the constituent quark question from another perspective, by
asking: how many quarks are there in a baryon?

The answer from spectroscopy would simply be: $N=3$. This is, of course, the
basis of the time-honoured Isgur-Karl model, and we have heard an updated
review on its further developments, successes and limitations by
S.~Capstick. From a different viewpoint, E.~Klempt reminded us of the important
role played by the color degree of freedom beyond just antisymmetrizing the
baryon wave functions.

When looking at deep-inelastic lepton-nucleon scattering the answer to the same
question would rather be: $N \to \infty$. This is the picture projected by the
HERA data and reviewed by R.~Yoshida. Counting quarks means, in this context,
taking the integral of $F_2 (x)/x$ over the Bjorken scaling variable $x$, or
equivalently, $\int dlnx \, F_2 (x) \to \infty$, given the observed behaviour
of the structure function $F_2 (x)$ at small $x$. (Of course one must recall
that small-$x$ physics, when viewed in the laboratory frame, involves
quark-antiquark and gluon fluctuations of the high-energy virtual photon which
extend over distances large compared to the proton size. The fact that $N \to \infty$ is therefore a statement about the interacting
photon-nucleon system at high energies, not about the isolated nucleon.)

Constituent quarks somehow interpolate between "three" and "infinity". It is
likely that their properties are linked to the rich, highly non-trivial
structure of the QCD vacuum. Y.~Simonov pointed out that constituent quark
masses and magnetic moments might have a direct relationship to the gluonic
string tension. D.~Diakonov emphasized the importance of instantons, large
fluctuations of the gluon field. In this approach, the constituent quark mass
at zero momentum is proportional to the average instanton size and inversely
proportional to the squared average instanton separation in the QCD
vacuum. With an instanton radius of about 1/3 fm and a separation of 1 fm,
compatible with instanton simulations in lattice QCD, the zero-momentum
constituent quark mass emerges at about 350 MeV, a remarkable result. The
"running" quark mass decreases rapidly at momentum scales larger than 1
GeV, just as it should. Lattice QCD studies of the Euclidean quark propagator
performed by the Adelaide group (D.~Leinweber et~al.) find a similar behaviour
of the momentum dependent quark mass in the Landau gauge (with the caveat to be
added that a constituent quark mass is not a gauge invariant quantity in QCD).

Confinement still persists as the primary challenge in all efforts trying to
understand the physical meaning of constituent quarks. Confinement is missing,
for example, in the instanton approach despite its apparent successes.

The nature of residual interactions between constituent quarks has been a much
debated theme. Spin-flavour correlations characteristic of Goldstone boson
exchange (as discussed e.~g. by L.~Glozman et~al.) have been held against
traditional one-gluon exchange descriptions. There is good reason to expect
that both kinds of forces act (amongst others) peacefully side by side. One
should not forget, after all, that a local interaction between two
color-currents generates, via Fierz transformation, all sorts of exchange
terms including color-singlet spin-flavour exchanges with meson quantum numbers.

\section{Baryon Resonances}
The research on baryon resonances has progressed enormously during the time
span between this and the last BARYONS conference. These new developments,
primarily driven by the experimental programmes of the CLAS detector at JLab,
at Mainz and Bonn and at the GRAAL and LEGS facilities, define a new level of
high-precision measurements. Out of the many highlights, only a few can be
discussed here, with no attempt to achieve even partial completeness.

An impressive example is the multipole analysis of neutral pion
electroproduction data, reported by V.~Burkert, which sets new standards for
the separation of magnetic dipole and electric quadrupole transition
formfactors in the delta resonance region at momentum transfers as high as 1
GeV. The search for missing resonances in the baryon spectrum is entering a new
era as well. CLAS data on two-pion electroproduction indicate signals in the
spectrum around center-of-mass energies of 1.7 GeV, which are not covered by
commonly used isobar models. Virtual  Compton scattering $(ep \to ep \gamma)$
with its minimal final state interaction also proves to be a valuable source of
information for systematic baryon resonance studies. At JLab's Hall A the first
measurement covering the entire region up to $W =$ 2 GeV has been performed
(reported by H.~Fonvieille) and displays strong resonance excitations. Very impressive results have also been presented
for resonance searches in electroproduction on the proton leading to
kaon-hyperon final states, again by a CLAS collaboration. To round off the
picture, T.~Nakano showed promising developments in kaon photoproduction at
SPRING-8 with focus on the $\Lambda (1405)$ and $\Lambda (1520)$, the detailed
nature of which is still a matter of debate between standard quark models and
coupled channels approaches.

New high quality data from GRAAL were reported by A.~D'Angelo. Following
previous pioneering measurements at Mainz and Bonn, eta meson photoproduction now covers
the region from near-threshold to 1.1 GeV in a single experiment, with stunning
accuracy and the quest for further explorations into a possible structure
around a photon energy of 1.05 GeV. Another very interesting case is the photon
beam asymmetry measured in omega meson production on the proton at photon
energies above 1 GeV. Purely diffractive $t$-channel exchange mechanisms would
not produce any asymmetry at all. Non-zero asymmetry signals measured with high precision
are therefore a distinct testing ground for $N^*$ resonance studies.

Polarisation observables are a key to the detailed analysis and understanding
of resonances in multipole amplitudes. Experiments with polarized photon beams
on polarized protons performed at Mainz and Bonn, and also recently at LEGS (as
summarized by A.~D'Angelo), are paving the way for an accurate examination of
the Gerasimov-Drell-Hearn sum rule. The relevant GDH integral when taken up to
about 1.8 GeV seems to pass beyond the canonical sum rule value by about 5--10
\%. A further extension of the double polarization measurements to higher
energies is therefore mandatory.

The theoretical state of the art on baryon resonance physics was reviewed by
T.~Sato. Various models exist to deal with the multipole amplitudes for photo-
and electroproduction, mostly based on effective Lagrangians with inclusion of
dominant baryon resonances. The situation can be summarized as follows. The
physics of the delta resonance is quite well under control. An exception is the
$N \to \Delta$ quadrupole transition amplitudes for which the detailed
understanding still needs to be improved. While models including pion cloud
effects are in good agreement with JLab data for $Q^2$ above 0.4 GeV$^2$,
A.~Bernstein pointed out in discussions that such calculations fail to
reproduce the accurate Bates data at $Q^2 \simeq$ 0.1 GeV$^2$ where the
theoretical treatment of pion cloud effects should actually be more reliable. In the second resonance
region it is important to improve on descriptions of channels with two pions in
the continuum. The theory of such $\pi \pi N$ three-body channels has not yet
reached a quantitatively satisfactory level. In the third and higher resonance
region there is still much work to do on the theoretical side, even conceptually,
concerning detailed resonance versus background analysis. Problems of this
sort, with quasi-bound states embedded in a multiparticle continuum, exist also
in nuclear physics, atomic physics and quantum optics, and it may be useful to borrow from the
reaction theory expertise developed in those fields.

Last not least, with the high productivity of several electron and photon
facilities now in operation and further upgrades in sight, one should not
forget the important information provided by experiments with hadron beams, as
pointed out by M.~Sadler. The picture will not be complete unless
electromagnetic and hadronic probes are systematically used in a complementary
way.

\section{Spin Structure}
Investigations of the nucleon's spin structure are entering a new phase. It is
now well established that quarks carry only a fraction, less than one third, of
the proton spin. The next challenges are: to extract the flavour decomposition
of those spin fractions, to isolate the gluon contribution to the total spin,
and to analyse transverse spin degrees of freedom which provide additional
independent pieces of information.

G.~van~der~Steenhoven gave an impressive   status report and an exciting outlook
into coming years. The HERMES experiment continues improving the data quality
for the polarized structure functions $g_1$ of the proton, the deuteron, and of
the neutron as deduced from $^{3}He$. The accuracy of the flavour decomposition
of quark spin fractions from the analysis of semi-inclusive deep-inelastic
scattering is expected to improve as well. The gluon contribution to the proton
spin will be explored with COMPASS, HERMES and RHIC, by photon-gluon fusion and
reconstruction of the produced charmed quark-antiquark pairs, and by measuring
tracks of high transverse momentum, with an expected accuracy at the 10 \%
level. A novel quantity of considerable current interest is the density of
transversely polarized quarks inside a transversely polarized proton, the
"transversity" distribution $h_1$. By its very nature, $h_1$ emphasizes more
prominently the valence quark aspects of spin structure, so it yields important
complementary information to the longitudinal quark spin distributions.

A further new dimension opens through Deep Virtual Compton Scattering (DVCS)
measurements and the extraction of Generalized Parton Distributions (summarized
by M.~Diehl). According to Ji's sum rule, DVCS has access to the total quark
angular momentum (its spin plus orbital angular momentum). The feasibility of
DVCS experiments has recently been demonstrated, in parallel, both by HERMES
and by the CLAS collaboration at JLab.

Another important issue is the $Q^2$ evolution of spin structure functions and
the influence of baryon resonances at smaller $Q^2$. JLab has taken a
leadership role in these projects, as reviewed by R.~de~Vita. The preliminary
data on the first moment of $g_1$ for the proton begin to provide the
systematics that is necessary to interpolate between the deep inelastic
scattering  region and the non-perturbative spin physics at lower $Q^2$, down
to the GDH sum rule at $Q^2 = 0$. A very interesting new result is the CLAS
measurement of $g_1 (x)$ for the proton at $Q \simeq$ 1 GeV. It demonstrates
the importance of the delta resonance (and possibly higher resonances) in the
nucleon spin structure problem, even at reasonably large $Q$. This behaviour
was predicted several years ago (Edelmann et~al.) and the theory has been
further developed in the meantime (Simula et~al.). To those who confidently
expected perturbative QCD to start already at $Q \sim$ 1 GeV, this came
unexpected. It is an interesting question now whether the basic idea of
quark-hadron duality (reviewed by S.~Jeschonnek at the conference), namely that
properly averaged hadronic observables can be described in terms of
perturbative QCD, is at work also for spin structure functions.

\section{Nucleon Form Factors and Polarizabilities}
A bright highlight in the exploration of nucleon electromagnetic form factors
(reviewed by E.~Brash) is the recent observation by a polarization transfer
measurement at JLab's Hall A that charge and magnetization distributions in the
proton are not proportional as was long thought. The ratio $\mu G_E /G_M$ drops
continuously down to almost one half at $Q^2 \simeq$ 4 GeV$^2$, whereas it was
previously assumed to be equal to one on the basis of simple universal dipole
parametrizations for those form factors. The JLab data have triggered a
re-examination of the Rosenbluth extraction of form factors from earlier cross
section measurements. Some relativistic constituent quark models and the cloudy
bag model describe the new situation properly, at least at a qualitative
level. Further thinking about the dynamics which governs the proton's
constituents and their wave functions will obviously be stimulated by these new
developments.

Parity violation in elastic electron scattering as a tool to investigate
admixtures of strange quark-antiquark pairs in the nucleon has been reviewed by
F.~Maas. Earlier predictions of large ground state matrix elements for strange
quark currents in the nucleon are not confirmed. The SAMPLE experiment at
Bates, the HAPPEX measurements at JLab and recent advances at Mainz all set
quite low limits for the strange vector current in the proton, considerably
lower than the 10 \% level expected from previous
estimates. This does not imply, of course, that strangeness is equally
unimportant in other types of currents. For example, the scalar density of
strange quarks in the nucleon might be considerably more pronounced. Renewed
interest in this question is triggered by recent claims that the sigma term of
the nucleon is substantially larger than its previously deduced value of $(45
\pm 8)$ MeV.

Electromagnetic and spin polarizabilities have always played an important part
in our understanding of nucleon structure. Virtual Compton Scattering (VCS,
reviewed by H.~Fonvieille) and the generalized polarizabilities deduced from
such experiments will be a rich source of new insights into the low-energy
dynamics of the nucleon's internal degrees of freedom. The accuracy reached in
the pioneering VCS measurements at Mainz is a major step forward, and together
with the VCS programmes at JLab and Bates, the progress in the field will be
significant. On the theoretical side, the analysis of VCS data is greatly
helped by sophisticated dispersion relation methods (B.~Pasquini et~al.) and by
effective field theory approaches (H.~Grie\ss hammer et~al.).

\section{Chiral Dynamics}
Low-energy QCD with light $u$- and $d$-quarks is realized in the form of an
effective field theory of pions, the Goldstone bosons of spontaneously broken
chiral symmetry, coupled to "heavy" sources such as baryons. The low-energy
expansion of this theory in small momenta or small quark masses is called
chiral perturbation theory (ChPT). Both its predictive capacity and its
inherent limitations were lucidly summarized by T.~Becher. Based on the
pioneering work by Gasser and Leutwyler and by Weinberg, ChPT has been a
remarkably successful theoretical tool, not only in its traditional domain,
threshold pion-pion scattering, but also in understanding the role of the
nucleon's pion cloud in pion-nucleon scattering and photoproduction processes.

One of those traditionally successful examples has been neutral pion
photoproduction on the proton close to threshold. A further challenging test
for ChPT is $\pi^0$ production induced by virtual photons and its systematic
$Q^2$ dependence. H.~Merkel reported on the very accurate Mainz measurements at
$Q^2 =$ 0.05 and 0.1 GeV$^2$, close to threshold, a region where the chiral
low-energy expansion is expected to work well. But apparently it doesn't! The
ChPT predictions of V.~Bernard et~al. systematically fail as one moves away
from the real-photon point to $Q^2 > 0$. (Admittedly, the MAID multipole
amplitudes have similar difficulties when confronted with these high-precision
data.) The issue raised by these findings is twofold: the data systematics
should be enlarged by measurements at several additional values of $Q^2$; and
the ChPT calculations will have to meet this challenge by moving to the next
higher order.

Rigorous ChPT is limited in its applicability to sometimes very small
convergence regimes when resonances are produced nearby. A prominent example is
the delta resonance. In the "official" power counting philosophy, its effects
are relegated to higher orders. But the prominent role of the $\Delta (1232)$
in the strong $M 1$ transition that governs the paramagnetic response of the
nucleon, an evident empirical fact, calls for a different kind of
book-keeping. In the large $N_C$ limit, the nucleon and the $\Delta$ are
degenerate: there is no particular reason to treat the baryon octet and the
            decuplet on a different basis in heavy-baryon ChPT. Extended
            versions of chiral effective field theory do in fact promote the
            delta to leading order in combined chiral and large-$N_C$
            expansions, a useful strategy. Such a scheme has been successfully
            employed, for example, in analysing the strong energy dependence of
            the generalized magnetic dipole polarizability of the nucleon (as
            reported by H.~Grie\ss hammer). It is also a key element in chiral
            extrapolations of nucleon properties deduced from lattice QCD
            (T.~Hemmert et~al., W.~Melnitchouk et~al.).

With inclusion of strangeness, non-perturbative features of meson-baryon
dynamics can be handled efficiently by combining the chiral SU(3) chiral
effective Lagrangian with coupled channels techniques familiar from nuclear
reaction theory. While sacrificing some of the puristic ChPT power counting
rules, the physics benefit from iterating important subclasses of driving
amplitudes to all orders is certainly rewarding. Most recent developments
(E.~Kolomeitsev, M.~Lutz), reported at this conference, now include the full
baryon decuplet in addition to the octet. The number of free parameters is
constrained by large-$N_C$ bookkeeping. The quantitative agreement of such
calculations with a large number of observables in a variety of meson-baryon
channels is quite remarkable.

Is the restoration of chiral symmetry (from its spontaneously broken
Nambu-Goldstone realization to the unbroken Wigner-Weyl mode) visible in the
high-mass sector of the baryon spectrum? This interesting question (raised by
L.~Glozman and T.~Cohen) alludes to possible parity dublets in the spectrum,
i.~e. degeneracies between positive and negative parity baryon resonances of
the same spin. Several such cases can be located in the mass region around 2
GeV, but there are several other examples which seem not to follow this
rule. The parity dublet criterion in its strict sense would apply to narrow
states, whereas most of the excited baryon states in question have large
widths. On the other hand, in the asymptotic continuum region, identical
spectral distributions of parity partners are a familiar phenomenon, as seen for
example in the large $s$ behaviour of vector and axial vector spectral
functions in the $\rho$ and $a_1$ channels. In any case, such discussions
emphasize again the need for more systematic explorations of high lying
excitations in the $N$ and $\Delta$ spectrum.

\section{Looking forward}
This conference has given us once again a very lively and convincing
demonstration that baryons (and their prototype, the nucleon) represent a most
fascinating QCD many-body problem. It exhibits all of the phenomena associated
with QCD, notably confinement, spontaneous chiral symmetry breaking and
asymptotic freedom. While the perturbative high-energy sector of QCD is quite
well understood, its non-perturbative features, the complexity of the QCD
vacuum and its relationship to confinement and spontaneous symmetry breaking
continue to be outstanding challenges. Last but not least, nuclear physics
needs ultimately to be understood in these terms, and the issue (alluded to in
G.~A.~Miller's talk) of how a nucleon changes its properties in a nuclear
environment (a QCD many-body problem intertwined with the nuclear many-body
problem) is as burning as ever.

One can say without exaggeration that the field is in a promising and
forward-looking situation. It is driven by three basic elements which are now
operating in a healthy balance: it is "data driven", with emphasis on high
precision and full utilization of polarization observables; it is "brain
driven", with advanced theoretical approaches constrained by symmetries of QCD;
and it is "computer driven", with future computational capacity and speed
progressing into the multi Teraflop regime.

Given the variety of running facilities at our disposal and upgrades in sight
(JLab, Bates, MAMI C, HERMES, COMPASS, RHIC-Spin, SPRING-8, ..., amongst
others), the perspectives for deepening our understanding of QCD phenomena are
certainly good. But one should also re-emphasize the importance of systematic
investigations using hadron beams. In particular, our knowledge about systems
with strange and charmed quarks is still underdeveloped.

\vspace{1cm}
I close with an expression of deep gratitude to Bernhard~Mecking, John~Domingo
and all members of the JLab team who have made this conference an exciting and
memorable event.
\end{document}